\documentclass[a4paper, 11pt]{article}

\usepackage[utf8]{inputenc} 

\usepackage{todonotes} 

\usepackage{lipsum}
\usepackage{adjustbox} 

\usepackage{xargs} 

\usepackage[backend=biber, sorting=none,
citestyle=numeric-comp, backref=true]{biblatex} 
\usepackage{authblk} 
\usepackage{braket} 
\usepackage{appendix} 
\usepackage{tabularx}
\usepackage{float}

\usepackage{graphicx} 
\graphicspath{ {images/} } 
\usepackage{caption}
\usepackage{subcaption}
\usepackage{mwe}

\usepackage{amssymb} 
\usepackage{mathtools} 
\usepackage{amsmath} 

\usepackage{braket} 
\usepackage{slashed} 
\usepackage{tensor} 
\usepackage{dsfont} 

\usepackage{float}

\usepackage{caption}
\usepackage{subcaption}

\usepackage{csvsimple}

\newcommand{\projectdir}{./}

\newcommand{\Romatre}
{Department of Mathematics and Physics,
 Roma Tre University, Rome, Italy.
}

\providecommand{\keywords}[1]
{
  \small
  \texttt{Keywords - } #1
}

\newcommand{\GeV}{\, \text{GeV}}

\newcommand{\EMsup}{\text{EM}}

\newcommand{\Csup}{\text{C}}
\newcommand{\PSEsup}{\Csup}

\newcommand{\sea}[1]{\textcolor{blue}{#1}}

\newcommand{\MASSsup}{\text{M}}

\usepackage{multicol}
\usepackage{hyperref}

\title{Leading Isospin Breaking effects in nucleon and $\Delta$ masses
\footnote{This work has been submitted to Suplemento de la Revista Mexicana de Física as a proceeding paper for the \href{https://indico.nucleares.unam.mx/event/1541/overview}{HADRON2021} international conference.
}
}
\author{Simone Romiti
\footnote{
\Romatre - Largo S. Leonardo Murialdo, 1, 00146 Rome RM, Italy
}
}
\date{July 2021}

\begin{document}

\maketitle

\begin{abstract}
  We present a lattice calculation of the leading corrections to the masses of nucleons and $\Delta$ resonances.
  These are obtained in QCD+QED at $1$st order in the Isospin Breaking parameters $\alpha_{EM}$, the electromagnetic coupling, and $\frac{{\hat{m}}_d - {\hat{m}}_u}{\Lambda_{QCD}}$, coming from the mass difference between $u$ and $d$ quarks.
\end{abstract}


\newpage

\section{Introduction}
\label{sec:introduction}

In this work the Leading Isospin Breaking Effects (LIBEs) in the spectrum of nucleons and $\Delta(1232)$ resonances are investigated using the RM123 method \cite{de2012isospin, tantalo2013isospin, PhysRevD.87.114505}.
%
%
%
Our calculation is done on the lattice,
using a mixed action approach with the twisted mass QCD (tmQCD) regularization over the $N_f=2+1+1$ European Twisted Mass Collaboration (ETMC) gauge configurations
\cite{carrasco2014up}.
%
%
A purely hadronic scheme is adopted in order to set the scale, tune the counterterms and extrapolate to the physical point.
%
%
The results found in this work are the following.
The uncertainties are only statistical and obtained using the jackknife resampling technique.
We obtain
$$
M_n - M_p
= \input{\projectdir QCDQED/Data/TeX/results/dM_np_feyn.dat}
\, \text{MeV}
\quad ,
$$
and the Isospin Breaking (IB) mass splittings in the $\Delta(1232)$ quadruplet (see tab. \eqref{tab:dM.Delta.results.MeV})
\footnote{
As we'll see, at Leading Order (LO) in IB the knowledge of only $2$ of them is sufficient to determine the others.
}.
\begin{table}[H]
\parbox{.45\linewidth}{
\centering
\begin{tabular}{cc}
 \input{\projectdir QCDQED/Data/TeX/Delta_masses/M_Deltas.dat}
\end{tabular}
\caption{Our results for the masses of the $4$ lightest $\Delta$ resonances.}
\label{tab:M.Delta.results.GeV}
}
\quad
\parbox{.45\linewidth}{
\centering
\begin{tabular}{cc}
 \input{\projectdir QCDQED/Data/TeX/Delta_masses/dM_Deltas.dat}
\end{tabular}
\caption{Our results for the $\Delta$(1232) mass splittings.}
\label{tab:dM.Delta.results.MeV}
}
\end{table}
We also get a prediction for the masses of nucleons,
\begin{align*}
  M_{n}
  &=
  \input{\projectdir QCDQED/Data/TeX/results/M_n.dat}
  \GeV{}
  \quad ,
  \\
  M_{p}
  &=
  \input{\projectdir QCDQED/Data/TeX/results/M_p.dat}
  \GeV{}
  \quad ,
\end{align*}
and of the $\Delta$ resonances (see tab. \eqref{tab:M.Delta.results.GeV}).

The paper is organized as follows.
In sec. \eqref{sec:LIBEs.lattice} we review the RM123 method and set our notation for the Isospin Breaking Effects (IBEs).
In sec. \eqref{sec:syst.tune.extr} we discuss the systematic effects, the tuning of counterterms and the extrapolations over the ensembles.
Finally in sec. \eqref{sec:conclusion} we give our conclusions.

\section{Leading IB effects on the lattice}
\label{sec:LIBEs.lattice}
At LO in IB we expand the path integral in the IB parameters $\Delta m_{ud}=(m_d-m_u)/2$ and $e^2$, taking into account $O(e^2)$ counterterms from QED diagrams divergences \cite{peskin1995introduction}
\footnote{
  Note that the fine structure constant $\hat{\alpha}_{EM}$ renormalizes at higher orders \cite{tantalo2013isospin},
  so that we can safely use the value
  \mbox{
  $
  \alpha_{EM}
  =e^2/(4\pi)
  =  1/137.035 999 084
  $
  }
  from \cite{10.1093/ptep/ptaa104}.
}.
In our twisted mass Lattice QCD (tmLQCD) approach \cite{PhysRevD.87.114505},
we have counterterms for both the physical and critical masses.
The Leading Isospin Breaking correction to the mass of an hadron $H$ is then:
\begin{equation}
  \Delta M_H =
  \left[
  e^2 \bar{\Delta}^{\text{\EMsup}}
  \, + \sum_{f} a \Delta m_{f}^{cr} \bar{\Delta}_f^{\Csup}
  \, + \sum_{f} a \Delta m_{f} \bar{\Delta}_f^{\MASSsup}
  \right]
  M_H
  \quad ,
\end{equation}
where the
$\bar{\Delta}^{\EMsup}$
and
$\bar{\Delta}_f^{x}$
\mbox{($x=\Csup, \MASSsup$)} (for a flavor $f$)
are the slopes induced by the couplings in front of them:
$\EMsup \to e^2$,
$\Csup \to$ (critical mass),
$\MASSsup \to$ (physical mass).
At $1$st order these are evaluated in isoQCD from the corrections $\bar{\Delta}^{x}$ in the euclidean correlators whose isoQCD ground state has mass $M_H^{(0)}$.
The mass slope's effective curve is \cite{PhysRevD.87.114505}
\footnote{
This formula holds in absence of backward signals,
namely for baryonic correlators with given parity \cite{sasaki2000n, Lee_1999},
while for mesons it gets slightly modified \cite{PhysRevD.87.114505}
}
:
\begin{equation}
  \label{eq:dM.from.dC.C0.no.back}
  \bar{\Delta}^{x} M_H(t)
  = - \partial_t
  \left[
  {\bar{\Delta}^{x} C_H(t)}/{C_H^{(0)}(t)}
  \right]
  \quad ,
\end{equation}
where $\partial_t f(t) = f(t+1)-f(t)$ (in lattice units).
In this work we extract the mass slopes fitting these curves to a constant in their plateaus.

We can then write the LIBEs in terms of Feynman diagrams reading \cite{PhysRevD.87.114505} for mesons and easily extending \cite{de2012isospin} for baryons.
For the latter we set a shorthand notation for the slopes
$\bar{\Delta}^{x} C_{H}^{(i)}$
\footnote{
The baryonic correlators $C_H$ are built from the interpolators of \cite{Alexandrou2014}.
}
,
where $x$ corresponds to the current insertion(s) and $i=1,2,3$ is the quark propagator index.
When $x=\MASSsup, \PSEsup$ we insert the scalar or pseudoscalar current respectively on the $i$-th quark leg, while $x=\text{self}$ comes from its electromagnetic self-energy.
When $x=\text{exch}$ we exchange a photon between the $2$ quarks different from the $i$-th one.
We define the ratios
\mbox{
${\mathcal{R}_H}_i^{x} = -\partial_t [{\bar{\Delta}^{x} C_{H}^{(i)}}/{C_{H}^{(0)}}]$
}
for $x \in \{\MASSsup, \PSEsup, \text{self}, \text{exch}\}$
and
\mbox{
${\mathcal{R}_H}_{i \sea{f} }^{\text{loop}} = -\partial_t [{\bar{\Delta}^{x} C_{H}^{(i \sea{f})}}/{C_{H}^{(0)}}]$
},
where the latter comes from the exchange of a photon between the $i$-th quark with a quark loop of flavor $\sea{f}$ (from the sea).

The LIBEs for nucleons then assume the following form:
\begin{equation}
  \label{eq:dMn.Feynman}
  \begin{aligned}
  &\Delta M_n
  =
  - \Delta m_u {\mathcal{R}_{N}}_1^{\MASSsup}
  - \Delta m_d {\mathcal{R}_{N}}_2^{\MASSsup}
  - \Delta m_d {\mathcal{R}_{N}}_3^{\MASSsup}
  \\
  &
  + \Delta m_u^{(cr)} {\mathcal{R}_{N}}_1^{\PSEsup}
  + \Delta m_d^{(cr)} {\mathcal{R}_{N}}_2^{\PSEsup}
  + \Delta m_d^{(cr)} {\mathcal{R}_{N}}_3^{\PSEsup}
  \\
  &
  + q_u^{2} {\mathcal{R}_{N}}_1^{\text{self}}
  + q_d^{2} {\mathcal{R}_{N}}_2^{\text{self}}
  + q_d^{2} {\mathcal{R}_{N}}_3^{\text{self}}
  + q_u q_d {\mathcal{R}_{N}}_{3}^{\text{exch}}
  + q_u q_d {\mathcal{R}_{N}}_{2}^{\text{exch}}
  + q_d^{2} {\mathcal{R}_{N}}_{1}^{\text{exch}}
  \\
  &
  +
  \sum_{\sea{f \in (sea)}}
  \sea{q_f}
  \left[
  q_u {\mathcal{R}_{N}}_{1 \sea{f} }^{\text{loop}}
  + q_d {\mathcal{R}_{N}}_{2 \sea{f} }^{\text{loop}}
  + q_d {\mathcal{R}_{N}}_{3 \sea{f} }^{\text{loop}}
  \right]
  \\
  &
  + [\text{isosymm. vac. pol. diag.}]
  \quad ,
\end{aligned}
\end{equation}
and $\Delta M_p$ is found via the exchange symmetry $u \leftrightarrow d$.
For the $\Delta$s we have:
\begin{equation}
  \label{eq:dM.Delta.pp.Feynman}
  \begin{aligned}
  &\Delta M_{\Delta^{++}}
  =
  - \Delta m_u
  [
  {\mathcal{R}_\Delta}_1^{\MASSsup}
  + {\mathcal{R}_\Delta}_2^{\MASSsup}
  + {\mathcal{R}_\Delta}_3^{\MASSsup}
  ]
  + \Delta m_u^{(cr)}
  [
  {\mathcal{R}_\Delta}_1^{\PSEsup}
  +{\mathcal{R}_\Delta}_2^{\PSEsup}
  +{\mathcal{R}_\Delta}_3^{\PSEsup}
  ]
  \\
  &
  + q_u^2
  [
  {\mathcal{R}_\Delta}_1^{\text{self}}
  + {\mathcal{R}_\Delta}_2^{\text{self}}
  + {\mathcal{R}_\Delta}_3^{\text{self}}
  + {\mathcal{R}_\Delta}_{3}^{\text{exch}}
  + {\mathcal{R}_\Delta}_{2}^{\text{exch}}
  + {\mathcal{R}_\Delta}_{1}^{\text{exch}}
  ]
  \\
  &
  +
  \sum_{\sea{f \in (sea)}}
  \sea{q_f} q_u
  \left[
  {\mathcal{R}_\Delta}_{1 \sea{f} }^{\text{loop}}
  + {\mathcal{R}_\Delta}_{2 \sea{f} }^{\text{loop}}
  + {\mathcal{R}_\Delta}_{3 \sea{f} }^{\text{loop}}
  \right]
  \\
  &
  + [\text{isosymm. vac. pol. diag.}]
  \quad ,
\end{aligned}
\end{equation}
\begin{equation}
  \label{eq:dM.Delta.p.Feynman}
  \begin{aligned}
  &3 \Delta M_{\Delta^{+}}
  =
  - \Delta m_d {\mathcal{R}_\Delta}_1^{\MASSsup}
  - \Delta m_u {\mathcal{R}_\Delta}_2^{\MASSsup}
  - \Delta m_u {\mathcal{R}_\Delta}_3^{\MASSsup}
  \\
  &
  + \Delta m_d^{(cr)} {\mathcal{R}_\Delta}_1^{\PSEsup}
  + \Delta m_u^{(cr)} {\mathcal{R}_\Delta}_2^{\PSEsup}
  + \Delta m_u^{(cr)} {\mathcal{R}_\Delta}_3^{\PSEsup}
  \\
  &
  + q_d^{2} {\mathcal{R}_\Delta}_1^{\text{self}}
  + q_u^{2} {\mathcal{R}_\Delta}_2^{\text{self}}
  + q_u^{2} {\mathcal{R}_\Delta}_3^{\text{self}}
  + q_d q_u {\mathcal{R}_\Delta}_{3}^{\text{exch}}
  + q_d q_u {\mathcal{R}_\Delta}_{2}^{\text{exch}}
  + q_u^{2} {\mathcal{R}_\Delta}_{1}^{\text{exch}}
  \\
  &
  +
  \sum_{\sea{f \in (sea)}}
  \sea{q_f}
  \left[
  q_d {\mathcal{R}_\Delta}_{1 \sea{f} }^{\text{loop}}
  + q_u {\mathcal{R}_\Delta}_{2 \sea{f} }^{\text{loop}}
  + q_u {\mathcal{R}_\Delta}_{3 \sea{f} }^{\text{loop}}
  \right]
  \\
  &
  + [\text{isosymm. vac. pol. diag.}]
  \\
  &+
  \{(d, u, u) \to (u, d, u)\}
  \,+\,
  \{(d, u, u) \to (u, u, d)\}
  \quad ,
\end{aligned}
  \end{equation}
$\Delta M_{\Delta^{-}}$ and $\Delta M_{\Delta^{0}}$ have the same form of $\Delta M_{\Delta^{++}}$ and $\Delta M_{\Delta^{+}}$ respectively, found via the flavor exchange $u \leftrightarrow d$.
It's easy to verify that at LO only $2$ of the $4$ $\Delta$ mass splittings are independent.
The IB correction to $M_{\Omega^{-}}$ is like $\Delta M_{\Delta^{++}}$, obtained replacing $u \to s$ (and the $\Delta$ interpolator with the $\Omega$'s).

\section{Systematics, tuning and extrapolations}
\label{sec:syst.tune.extr}
In this work we neglect the disconnected isosymmetric vacuum polarization diagrams \cite{ruuskanen1978unitarity}.
We also work in the electroquenched approximation \cite{frezzotti2016sea},
so that all the diagrams with photons attached to quark loops vanish.
We introduce QED in a non-compact way \cite{duncan1996electromagnetic},
with the QED$_{\text{L}}$ regularization for the photon propagaor \cite{giusti2017leading}.
The universal QED Finite Volume Effects (FVEs) in the hadronic spectrum \cite{10.1143/PTP.120.413, davoudi2014finite, FODOR2016245, borsanyi2015ab} are corrected for each ensemble,
leaving only the structure-dependent FVEs starting from $O(1/L^3)$.

In tmQCD the presence of IB leads to counterterms to both the critical and physical masses.
Analogously to \cite{PhysRevD.87.114505}, the former are tuned using the PCAC Ward Identity
requiring to preserve the maximal twist in isoQCD \cite{baron2010light} also at $O(e^2)$.
For an observable $O$, the quarks masses physical point in isoQCD and QCD+QED is defined by the ratios
$
r_s =
 [
 {
 2(M_{K^+}^2 + M_{K^0}^2)
 -(M_{\pi^+}^2+M_{\pi^0}^2)
 }
 ]
 /{2 M_{\Omega^{-}}^2}
$,
$
r_\ell  =
{(M_{\pi^+}^2+M_{\pi^0}^2)}/{2 M_{\Omega^{-}}^2}
$
and
$
 r_p     =
 {M_{K^+}^2}/{M_{\Omega^{-}}^2}
$
,
requiring them to match their experimental values.
As a consequence, at the physical point their total IB corrections vanishes.
We impose the latter condition at fixed ensemble in order to tune the counterterms $a \Delta m_f$.
This allows to evaluate (in the full theory and for each ensemble) any observable $O$,
whose physical point is reached by contruction after the aforementioned extrapolation.
The latter is done in separate steps,
on the slice $r_s=r_s^{(exp)}$ of the hyper-surface $O(r_s, r_\ell, L, a)$.

We extrapolate to the physical point, $L\to\infty$, and $a\to 0$
\footnote{
In this work the lattice spacings $a_{\beta(i)}$ ($\beta=1.90, 1.95, 2.10$, see \cite{carrasco2014up})are set by the ${\Omega^{-}}$ mass,
extrapolating $a M_{\Omega^{-}}$ among the ensembles with the polynomial ansatz:
\begin{equation}
 \label{eq:aMOmega.ansatz.Ll.QCDQED}
 (aM_\Omega)_i (L, r_\ell) \, = \,
 a_{\beta(i)} \, M_\Omega^{(exp)}
 \left[
  1 + \,
  c_{L} \frac{\alpha_{EM}}{L^3} + \,
  c_\ell \, r_\ell \, + \,
  c_{\ell}^{(2)} \, r_\ell^2
  \right]
 \quad .
\end{equation}
and setting the extrapolated values equal to $a_{\beta(i)} M_{\Omega^{-}}^{\text{exp.}}$.
The coefficients $a_{\beta(i)}$, $c_L$, ... are free parameters of the fit.
}
with global fits among the ensembles using phenomenological ans{\"a}tze inspired by LO ChPT \cite{Leutwyler:2012, alexandrou2009low, bernard2008chiral}.
The previously mentioned masses $M_i$ are fitted among the ensembles using the following functional forms:
\begin{equation}
  \begin{aligned}
    M_{i}(L, r_\ell, a) =
    A_{i} \,
    &
    \left[
    1
    + \alpha_{EM} \frac{c_3^{(i)}}{{L}^3}
    + c_a^{(i)}  a^2
    + c_\ell^{(i)}  r_\ell
    + c_{3/2}^{(i)} r_\ell^{3/2}
    \right]
    \quad ,
 \end{aligned}
\end{equation}
while for the IB mass splittings $\Delta M_i$ we use a simple polynomial ansatz:
\begin{equation}
  \begin{aligned}
    \Delta M_{i}(L, r_\ell, a) =
    D_{i} \,
    &
    \left[
    1
    + \alpha_{EM} \frac{c_3^{(i)}}{{L}^3}
    + d_a^{(i)}  a^2
    + d_\ell^{(i)}  r_\ell
    \right]
    \quad ,
 \end{aligned}
\end{equation}
The coefficients $A_i$, $c_a^{(i)}, ...$ and $D_i$, $d_i^{(a)}$, ... are left as free parameters of the fits.
Given the maximal twist (and hence the $O(a)$ improvement), discretization effects start at $O(a^2)$,
while the $\sim 1/L^3$ term accounts for the residual structure-dependent QED FVEs in the $a \Delta m_f$ and the masses themselves.
Higher orders in $1/L$ and QCD FVEs are found to be numerically negligible at our level of precision.

\section{Conclusion}
\label{sec:conclusion}
In this work we've computed on the lattice the LIBEs in the spectrum of mesons and baryons,
getting a prediction for the masses and IB mass splittings of nucleons and the $\Delta(1232)$ resonances.
In particular, we note that the full spectrum of the $\Delta(1232)$ quadruplet is not completely determined experimentally yet, motivating its investigation.
The values we found are compatible within at most $1.5\sigma$ with the experimental predictions.
This is so despite the approximations introduced in the calculation, indicating that at our level of precision the neglected diagrams are physically suppressed as expected.
Their neglection introduce nevertheless systematic effects which can be known only by direct evaluation,
and which we aim to include in a future work.
\newpage
\printbibliography
\addcontentsline{toc}{chapter}{Bibliography}

\end{document}